\documentstyle[epsf]{article}
\makeatletter
\def\epsfile#1{\def\@psfile{}\parse@ps@parms{#1}}
\def\parse@ps@parms#1{%
  \@for\@epsfile:=#1\do{\expandafter\@setparms\@epsfile,}%
  \epsfbox{\@psfile}}
\def\@setparms#1=#2,{\@nameuse{@setps#1}{#2}}
\def\@setpsfile#1{\def\@psfile{#1}}
\def\@setpsheight#1{\epsfxsize=0pt\epsfysize=#1}
\def\@setpswidth#1{\epsfxsize=#1\epsfysize=0pt}
\def\@setpsscale#1{\def\epsfsize##1##2{#1##1}}
\makeatother
\begin{document}
\begin{centering}
\LARGE {\bf Modeling of ions energy distribution profile of
electronegative plasma discharges with 
an efficient Monte Carlo simulator}
\\  \vspace{.75in}
\Large {M. Ardehali}
\\  \vspace{.4in}
\large {Silicon Systems Research Laboratories,
NEC Corporation, 
Sagamihara,
Kanagawa 229
Japan}
\\ \vspace{.5in}
\end{centering}

\begin{abstract}
The crucial role that Ions Energy Distribution Function
(IEDF) at the electrodes plays in
plasma processing of semiconductor materials
demands that
this quantity be predicted with high accuracy and with low
noise levels
in any plasma simulator.
In this work, 
an efficient
Particle-in-cell/Monte-Carlo (PIC/MC) simulator is developed to
model IEDF at the electrodes of 
electronegative plasma discharges.
The simulator uses
an effective method to
increase the number of MC particles in 
regions of low particle density by
splitting the particles and
by adjusting their statistical weight.
This statistical enhancement technique, which does not
require interprocessor communication, is particularly suitable for
parallel processing.
The simulator is used to model an electronegative rf discharge at
a pressure of 25 mTorr. 
The IEDF obtained from this simulator has good statistics
with low noise levels, whereas the IEDF calculated by standard
PIC/MC simulator is jammed with stochastic noise.
\\  \vspace{.1in}
\end{abstract}
\pagebreak

\begin{center}
\Large  {\bf I. Introduction}
\end{center}

Radio frequency glow discharges are widely used in microelectronics for
deposition and etching thin solid films and 
for Ultra Large Scale Integrated (ULSI) circuit 
fabrication \cite {1}.
The understanding of these discharges is of considerable importance for
a better control and optimization of semiconductor processing.
Unfortunately, these
discharges are complex systems that
are very difficult to analyze [2-4]. In the last few years,
Particle-In-Cell/Monte-Carlo (PIC/MC) simulation has been used
successfully to model rf discharges.
PIC/MC technique combines the microscopic transport of electrons and
ions with the self consistent electric field. The technique is very
attractive because it can handle the nonlocal effects which are
dominant at low
pressures without making any {\em ad hoc}
assumptions [5-6]. Furthermore, the PIC/MC technique is capable of
predicting ions energy distribution function
at electrode surfaces.

Unfortunately, PIC/MC simulation of RIE 
discharges suffers from significant noise which is
inherently associated with regions of low particle density.
In particular, because the particle
density in the sheath is many orders of magnitude smaller than
in the bulk, the MC noise inevitably
associated with the
sheath is 
considerably larger than the noise associated with the bulk.
Several techniques have been proposed to improve the 
sampling within the sheath [7-9].
However, even with
the improved sampling techniques,
the present PIC/MC simulators suffer from
intensive computational requirements and from
excessive noise 
associated with the sheath.
Serious problems arise from the noise within the sheath
and they may threaten the reliability of the simulation data.

\begin{center}
\Large  {\bf II. Details of simulation}
\end{center}

Since the motion of electrons and ions in rf discharges
under the influence of 
self-consistent electric field bears some resemblance to the motion
of electrons and holes in semiconductor devices, 
the advanced
techniques that have been developed for
device simulation 
[10-12] may
also be useful for plasma simulation.
For example, for hot electron injection in MOSFETs \cite{13}, 
only one out of $10^{10}$ channel 
electrons makes a contribution to the gate current. A
straight forward MC simulation is extremely expensive, requiring the
simulation of billions of cold MC particles. To reduce the 
simulation cost of 
these problems,
Phillips and Price \cite {10}
divided the
domain of the electron states {\em S} 
into a rare part {\em R}
and a complementary common part {\em C}.
When a particle enters {\em R} from {\em C},
its state is stored.
The rare portion of the particle history is then repeated for
a fixed number of $N$ times starting from the same initial condition.
After $N$ repetitions, the MC simulation continues in the {\em C} domain
using only one of the $N$ rare trajectories. Obviously, in
calculating the average values, a weight of $1/N$
is given to the results in {\em R}.

Motivated by their work, we use a similar technique to increase the
number of MC particles in the sheath by splitting the particles
in regions of low particle density.
There are,
however, some fundamental differences between this work and that of 
[10-12]. In particular, in the work of Phillips and Price, the energy 
components of the particles are weighed differently, whereas in the
present work,
as is shown in part III,
the spatial components of the particles are weighed differently.

There are also some important differences between this work and
the previous plasma simulation
works that increase the number of MC particles within
the sheath [14-16]. For example,
in the work of Goto {\em et. al.} \cite{16}, the discharge is
divided into many slabs  and every slab consists of $M_s$ sub-slabs.
Each sub-slab in turn is divided into $M_f$ fine-sub-slabs in which the 
generation and loss of particles are counted.
The number of particles in every fine-sub-slab
is set to a {\em standard} number $N_s$. When the 
simulated particle number of a slab exceeds $1.4M_sM_fN_s$, the 
number of particles
of all fine-sub-slabs in the slab is counted. If
the particle number of a fine-sub-slab is
greater than $N_s$, then it is
reduced to $N_s$.
Thus in their work, as well as in Refs. [14-15],
an attempt is made to set the number of particles
in every fine-sub-slab or in every cell equal to a 
constant number $N_s$. In contrast, in this
work, the discharge is not divided into slabs, sub-slabs, and
fine-sub-slabs, and no attempt is made to set the number of 
particles in every cell
or in every fine-sub-slab equal to a standard number. 
In the present work,
a suitable
weight $W_i$ is associated with 
each cell $i$
which is automatically updated with every time step, and 
unlike Refs. [14-16], the particles
are split when
the particle's weight is significantly larger than the
weight associated
with the cell.
In this respect the present work is similar to
the splitting technique used in the device simulation [10-12].

\begin{center}
\Large  {\bf III. Simulation of electrongative discharge with
SPLIT-PIC/MC technique}
\end{center}

A question of 
fundamental interest is whether the SPLIT technique, which was
applied to electropositive discharges \cite{17}, can also be 
used to improve the statistics of  electronegative simulators.
To address this question, we developed a simulator to model 
electronegative gases [18-20]
using SPLIT technique.
The present simulator uses an effective
method to increase of the number of
MC particles in regions of low particle density
by splitting the particles and by
adjusting their statistical weight.
The computation time for splitting the MC particles is
very small and in general is
less than $5\%$
of the total calculation time.
Using this splitting technique, the number of MC particles within
the sheath becomes of the same order as in the bulk. In contrast,
in the standard PIC/MC simulation, 
even though the
sheath is the most important region in the discharge and has a profound
influence on the properties of the bounded plasma,
the number of MC particles in the sheath is many
orders of magnitude smaller than in the bulk.

Very briefly,
the present SPLIT- PIC/MC simulator
consists of the following steps:
\\
(1) Each cell $i$ contains a suitable
weight $W_i$ which is automatically updated with time.
\\
(2) The instantaneous locations of MC particles
representing ions
and electrons are 
interpolated to the grid points to obtain the charge density.
\\
(3) Poisson's equation on a
spatially discretized mesh is solved to obtain the electric field.
\\
(4) The electric field is interpolated from the grid points to the
location of MC particles.
\\
(5) The equations of motion under the local and
instantaneous electric field are integrated.
\\
(6) When ion
$j$ enters cell $i$, its weight $m_j$ is compared
with $W_i$. If the particle's
weight is much heavier than $W_i$, then
it splits into two sub-particles, each with
weight $\frac{\displaystyle m_j}{\displaystyle 2}$.
\\
(7) Random numbers (Monte Carlos technique)
and collision cross sections are used to
determine the probability that each particle suffers
a collision.

It is worth noting that the present SPLIT-PIC/MC simulator
is particularly suitable for parallelization and vectorization. The
SPLIT method does not
use the distribution of particle weights in a cell to adjust 
the number of MC particles to a desired population.
Since the SPLIT technique does not require fetching 
particle data handled by other processors,
there is no need for interprocessor communication,
and complete parallelization 
with respect to particles is achieved.
The numerical results presented here were all
performed on an NEC
parallel machine Cenju-3 ($16$ processor)
which has a VR4400SC (75 MHz) processor and
$64$MB local memory in each processor \cite{21}.

To examine the effectiveness of SPLIT technique,
we have modeled an electronegative rf plasma at
a pressure of $25$ mTorr both with the SPLIT-PIC/MC simulator and
with the standard PIC/MC simulator \cite {22}.
The simulator is based on a Chlorine-like gas.
The processes considered are
ionization leading to positive
ion formation,
dissociative electron attachment
leading to negative ion formation, 
and positive ion/negative ion recombination leading
to positive and negative ions removal.
These processes are represented as follows:
\begin{eqnarray}
Cl_2 + e \rightarrow Cl_{2}^{+} + 2e \\ 
Cl_2 + e \rightarrow Cl^- + Cl\\ 
Cl_{2}^{+} + Cl^{-} \rightarrow Cl_{2} + Cl
\end{eqnarray}
where $Cl_{2}^{+}$ and $Cl^{-}$ are the main positive and
negative ions respectively.
In the simulation,
the left electrode is driven 
with a voltage $V_{rf}(t)= V_{rf} sin\omega t$, where
$V_{rf}= 300$ Volts,
the applied frequency
is $\frac{\displaystyle\omega}{\displaystyle2\pi}=10$ MHz,
and the discharge length is $10$ cm with perfectly
absorbing electrodes. 
The time step $\delta t=0.1 \, ns$ which
is small enough to resolve the electron
plasma frequency.

Since in this work we are primarily interested in testing the
effectiveness and usefulness
of SPLIT-PIC/MC simulator, we have intentionally
kept the model of the plasma simple.
The electron-neutral processes consist of elastic scattering and
ionization.
The total electron-neutral 
scattering cross section $\sigma_{total}(v)$ is given by
$\sigma_{total}(v)=\frac{\displaystyle K_{total}}{\displaystyle v}$,
where $v$ is the electron
velocity and the rate constant $K_{total}$
is $5 \times 10^{-8}$ $cm^3/s$ \cite{5}.
Ionizing collisions occur if the electron energy is larger than
$16$ eV. 
An ionizing collision [Eq. (1)] is modeled by loading a new electron
and ion at the position of the ionizing electron. The kinetic
energy after ionizing  collision is partitioned between the two
electrons with equal probability.
The cross section for ion-neutral charge-exchange
collision is 
$\sigma_{ce}(v) = 1\times 10^{-15} cm^2$ \cite{5}.
After charge exchange collision, the ion is
scattered isotropically, with its energy being set
at the background temperature.
Since the rate constant for attachment $K_{at}$ is 
constant \cite{18},
the cross section for attachment [Eq. $(2)$] is modeled as
$\sigma_{at}(v)=\frac{\displaystyle K_{at}}{\displaystyle v}$ 
where $K_{at}=1.8 \times 10^{-10}$ $cm^3/s$ \cite{18}. The 
attachment reaction is modeled by removing 
the electron from 
the system and by loading a negative ion at the position of the
attaching electron. The kinetic energy of the generated 
negative ion is set
at the background temperature. Finally because the 
rate constant for recombination $K_{at}$ is also constant,
the cross section for recombination [Eq. $(3)$] is modeled as
$\sigma_{rec}(v)=\frac{\displaystyle K_{rec}}{\displaystyle v_{ion}}$,
where $K_{rec} =5 \times 10^{-8}$ $cm^3/s$ \cite{18},
and $v_{ion}$ is the velocity of the recombining ion.
The recombination
reaction is modeled by removing positive and negative ions from 
the system.
Simulation results
have shown that the
splitting technique leads
to reduction of noise within the sheath no matter what the
detailed forms of the
cross sections are.

\begin{center}
\Large  {\bf IV. Results and discussion}
\end{center}

The simulation is started by loading a few hundred {\em quiet}
particles 
uniformly between the electrodes. Initially the velocity of
all particles, negative ions, positive ions and electrons, is set
to zero and the density of positive ions
is set equal to the density 
of negative ions and electrons. Since the mass of electrons is
much smaller than the mass of ions, 
electrons rapidly leave the system and
sheaths develop at the boundaries.
The density of electrons and ions continues to rise by ionization until 
steady state is reached where
the ionization rate becomes equal to the total loss rate.
At steady state the number of electrons and ions
approaches a constant value.
With SPLIT-PIC/MC technique, the simulator approaches steady state
after approximately $900$ cycles, whereas
with the standard PIC/MC simulator, more than
$2000$ cycles
are required for the simulator to
arrive at steady state.

To clearly demonstrate the effectiveness of the SPLIT technique,
we ran
SPLIT-PIC/MC simulator for
approximately $1000$ cycles, while we ran
the standard PIC/MC simulator for
approximately $3000$ cycles. Since the process of splitting 
particles takes less than $5\%$ of the calculation time,
the total computation time for the  
SPLIT simulator 
is approximately three times smaller than for the
standard simulator.
As shown below,
even with $66\%$ smaller computation time,
the simulation results obtained from SPLIT-PIC/MC simulator are
more accurate and have 
lower noise levels than the results obtained 
from the standard PIC/MC simulator.

Figures 1 (a) and 1 (b) show the number of MC particles
for positive ions
obtained from the SPLIT-PIC/MC simulator and from the standard
PIC/MC simulator (in relative units).
The important point to notice is 
that for SPLIT-PIC/MC simulator,
the number of MC particles  in the sheath
is of the same order of 
magnitude as in the bulk.
In contrast, for the
standard 
PIC/MC simulator, the 
number of MC particles
in the sheath is many orders of magnitudes smaller
than in the bulk.

Figures 2 (a) and 2 (b) show the instantaneous
(not averaged over several period) positive ion density
in log scale
obtained from the SPLIT-PIC/MC simulator and from 
the standard PIC/MC simulator. Note that
the massive ions do not respond to the rapid variations of the 
electric field and hence the ion density profile is nearly
time independent.
Figs. 2 (a) and 2 (b)  clearly indicate that
SPLIT
technique leads to significant reduction of MC noise within the 
sheath.
Figure 2 (c) shows the positive
ion density profile within the sheath in linear
scale
simulated by the standard PIC/MC technique (dashed) and by
the SPLIT-PIC/MC technique (line).
Comparing the two cases, it can be seen that
the PIC/MC profile is zero within several cells, whereas the
SPLIT-PIC/MC  profile is
never zero within any cells and has a much smaller noise level.

Figure 3 shows the instantaneous profile of 
negative ion density within the discharge simulated by SPLIT-PIC/MC
technique. 
The negative ions
are confined to the bulk of the discharge and are
completely excluded from the sheath by the
large sheath electric field.

Figures 4 (a) and 4 (b) show the 
instantaneous positive ion phase-space simulated by 
the SPLIT-PIC/MC technique and by the standard PIC/MC technique. 
These figures  clearly show that the ions are 
accelerated by the sheath electric field. The ions that suffer several
collisions arrive at
the electrode with very small energies. In contrast, the
ions that do not suffer any
collisions arrive at the electrode by the average sheath 
potential. The important point about these figures is that
the SPLIT-PIC/MC simulator
dramatically increases the number of ions within the sheath,
hence leading to significant reduction of MC noise in regions of 
low particle density.

As another example of the effectiveness of SPLIT technique, the
variation of the electric field within the 
discharge at phase $\frac{\displaystyle 3\pi}{\displaystyle 2}$
is simulated both by the
SPLIT-PIC/MC [Fig. 5 (a)]  technique and by the
standard PIC/MC [Fig. 5 (b)] technique.
Due to disparity between
electrons and ions masses, electrons rapidly leave the system and
large electric fields develop near the electrodes.
The electric field in the bulk of the discharge, however,
is much smaller than 
in the sheath,
indicating that the net charge density within the bulk
is very small.
The large electric field at the electrode boundaries
effectively pushes the 
negative ions out of the sheath (see Fig. 3).
Figure 5 (a) clearly demonstrates that
SPLIT-PIC/MC technique, 
by increasing the number of MC particles within the sheath,
dramatically
reduces the noise associated with
the sheath electric field
[it should be emphasized that these
figures {\em are not} averaged
over many periods, rather they represent the snapshot or 
the instantaneous electric field.
Furthermore, the
computation time for Fig. 5 (b) is three times larger
than for Fig. 5 (a)].

To demonstrate that the present
SPLIT-PIC/MC simulator can predict the physical 
properties of the electronegative discharges,
in Figs. 6 (a) and 6 (b), we present positive
and negative ion current densities 
averaged over 50 periods and 
at four times during the rf cycle.
(In contrast to Figs. 2-5, Figs. 6(a) and 6 (b) are averaged over
many periods and are not instantaneous).
Note that
the negative ion current density is almost one  
order of magnitude smaller than the positive ion current density.
Furthermore, since negative ions are excluded from the sheath, the
negative ion flux within the sheath is almost zero. In contrast, 
positive ions are accelerated by the sheath electric field and
positive ion flux increases sharply within the sheath.

Figures 2-5 have clearly shown the effectiveness of SPLIT-PIC/MC
simulator for improving the statistics of
ion density and the discharge electric 
field. However, from
the point of view of semiconductor processing, the most
important characteristic of RIE discharges is the 
Ions Energy Distribution
Function (IEDF)
at the 
surface of the electrodes [23, 24].
Thus a question of fundamental interest is whether the SPLIT
technique can improve the statistics of IEDF at
the electrodes
(the question of the effectiveness of SPLIT technique on IEDFs at 
the electrodes was not
considered in \cite{17}).
To answer this question, we simulated the IEDF both by the
standard technique and by the SPLIT technique.
The IEDF obtained from
the SPLIT-PIC/MC simulator and
from the standard PIC/MC simulator are shown in
Figs 7 (a) and 7 (b).
Figure 7 (a) is obtained after running SPLIT-PIC/MC simulator for
950 periods, whereas 
Fig. 7 (b) is calculated after running PIC/MC 
simulator for 3000 periods; however, in 
both Figs. 7 (a) and 7 (b), the IEDFs are time averaged over 50 periods.
The IEDF at the
electrode consists of a few spikes that are 
superimposed on a collisional background. Although most ions have 
many collisions in the sheath and 
hence create the collisional background
profile, a few ions have only several charge-exchange collisions, 
creating the spikes in the profile. Comparing Figs. 7 (a) and 
7 (b), it can be seen that with the standard PIC/MC simulator,
the 
statistics of IEDF
is very poor, jammed with stochastic noise; whereas with the
SPLIT-PIC/MC simulator, the statistics of IEDF is significantly 
enhanced and has a much lower noise level.
To further demonstrate the effectiveness of the SPLIT technique, in 
Figs. 7 (c) and 7 (d), we present the IEDF at the electrode
simulated by the 
standard PIC/MC technique and time averaged over $150$ and
$600$ periods respectively. Note that
Fig. 7 (a) which
is obtained from SPLIT-PIC/MC simulator and is
time averaged over only 50 period
is smoother than Fig. 7 (d) which is
obtained from  the standard PIC/MC simulator and is
time averaged over $600$ periods.
These figures clearly demonstrate that SPLIT technique leads to
dramatic reduction of stochastic noise in IEDF at the electrodes.

\begin{center}
\Large  {\bf V. Conclusion}
\end{center}

A stable algorithm which is highly suitable for parallelization
and vectorization is developed to simulate electronegative
rf discharges. The algorithm 
splits Monte-Carlo particles in
regions of low particle density and hence 
improves the sampling and reduces the noise
inherently associated with the sheath.
The process of splitting
the particles takes only a very small fraction
(less than $5\%$) of
the total computation time.
The most attractive feature of SPLIT technique is that it can 
dramatically reduce the noise associated with the IEDF at the electrode.
In particular, 
the IEDF at the electrode obtained from
SPLIT-PIC/MC simulator
and averaged over $50$ periods has a 
better statistics and lower noise level
than the distribution function obtained from 
the standard PIC/MC simulator
and averaged over 600 periods.
The SPLIT-PIC/MC simulator shows great promise for modeling
any kind of bounded plasmas where sheaths inevitably develop,
and for calculating IEDF 
at the electrodes of
more complex discharges
(such as ECR or ICP) in two or three dimensions.

\pagebreak

\pagebreak
{\large
Figure Captions}
\\
{\em \bf Important Note:} All figures should be
considered from left to right and top to bottom.
\\
Figures 1 (a), 1 (b),
are on the same page.
\\
Figures 2 (a), 2 (b), and
2 (c) are on the same page.
\\
Figure 3 is on one page.
\\
Figures 4 (a) and 
4 (b) are on the same page.
\\
Figures 5 (a) and
5 (b) are on the same page.
\\
Figures 6 (a) and
6 (b) are on the same page.
\\
Figures 7 (a) and
7 (b) are on the same page.
\\
Figures 7 (c) and
7 (d) are on the same page.
\\
{\bf Fig. 1 (a)} - Number of MC particles for ions
obtained from  SPLIT-PIC/MC simulator. Simulation conditions:
$V_{rf}= 300$ Volts,
$\frac{\displaystyle\omega}{\displaystyle2\pi}=10$,
pressure 25 mTorr. These
conditions apply to all figures. In Figs. [1-6],
$x=0$ (left electrode) is the powered electrode and 
$x=10$ (right electrode) is the grounded electrode.
\\
{\bf Fig. 1 (b)} - Number of MC particles for ions obtained from 
the standard PIC/MC simulator.
\\
{\bf Fig. 2 (a)} - Instantaneous ion density 
obtained from SPLIT-PIC/MC simulator in log scale.
\\
{\bf Fig. 2 (b)} - Instantaneous ion 
density obtained from the standard PIC/MC simulator in log scale.
\\
{\bf Fig.  2 (c)} - Instantaneous ion
density in the sheath from PIC/MC simulator (dashed)
and from SPLIT-PIC/MC simulator (line). 
\\
{\bf Fig.  3} - Instantaneous negative ion
density within the discharge
obtained from
SPLIT-PIC/MC technique in log scale. 
\\
{\bf Fig.  4 (a)} - Ion phase space diagram obtained from 
SPLIT-PIC/MC simulator.
\\
{\bf Fig.  4 (b)} - Ion phase space diagram obtained from 
the standard PIC/MC simulator.
\\
{\bf Fig. 5 (a)} - Instantaneous electric field at phase
$\frac{\displaystyle 3 \pi}{\displaystyle 2}$
obtained from SPLIT-PIC/MC simulator.
\\
{\bf Fig. 5 (b)} - Instantaneous electric field at phase
$\frac{\displaystyle 3 \pi}{\displaystyle 2}$
obtained from PIC/MC simulator.
\\
{\bf Fig. 6 (a)} - Positive ion current density averaged over 50 
periods at 4 times during the rf cycle. For
Figs. 6 (a) and  6 (b)
$\frac{\displaystyle \omega t}{\displaystyle 2 \pi}=0$ (solid line),
$\frac{\displaystyle \omega t}{\displaystyle 2 \pi}=0.25$
(dotted line),
$\frac{\displaystyle \omega t}{\displaystyle 2 \pi}=0.5$
(dashed-dashed line),
$\frac{\displaystyle \omega t}{\displaystyle 2 \pi}=0.75$ (dashed line).
\\
{\bf Fig. 6 (b)} - Same as Fig. 6 (a) but for 
negative ion current density.
\\
{\bf Fig. 7 (a)} - Ions energy distribuion profile at the 
electrode obtained from
SPLIT-PIC/MC simulator after running the simulator for
$1000$ periods and averaging over
$50$ periods. In Figs. [7 (a), 7 (b), 7 (c), and 7 (d)], 
the horizontal axis represents the
energy of ion striking the electrode, and the vertical axis represents
the distribution of ions energy.
\\
{\bf Fig. 7 (b)} - Ions energy distribuion profile at the 
electrode obtained from
PIC/MC simulator after running the simulator for
$3000$ periods and averaging over
$50$ periods.
\\
{\bf Fig. 7 (c)} - Ions energy distribuion profile at the 
electrode obtained from
PIC/MC simulator after running the simulator for
$3000$ periods and averaging over
$150$ periods.
\\
{\bf Fig. 7 (d)} - Ions energy distribuion profile at the 
electrode obtained from
PIC/MC simulator after running the simulator for
$3000$ periods and averaging over
$600$ periods.

\pagebreak
\epsfile{file=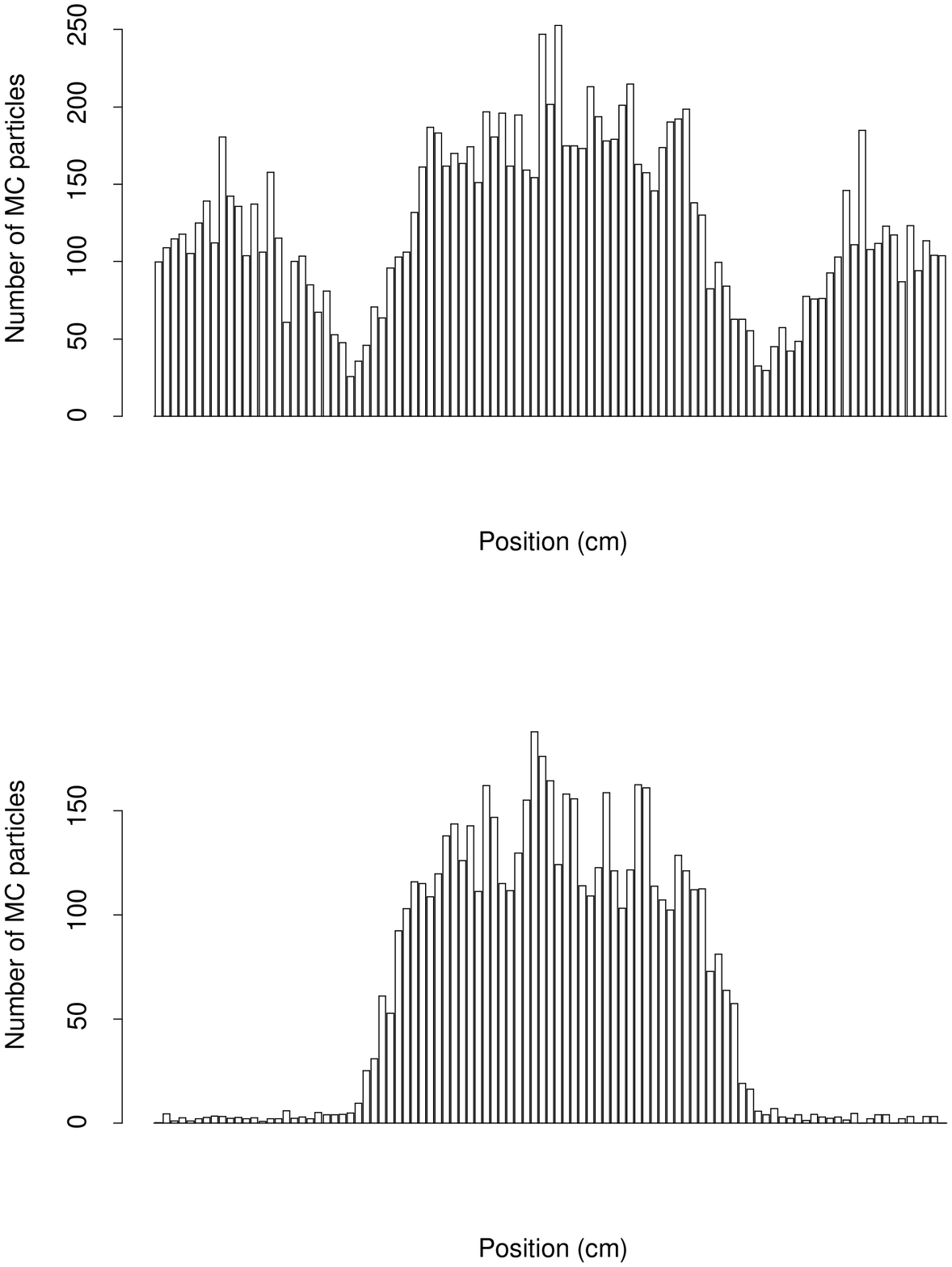,width=\textwidth}
\epsfile{file=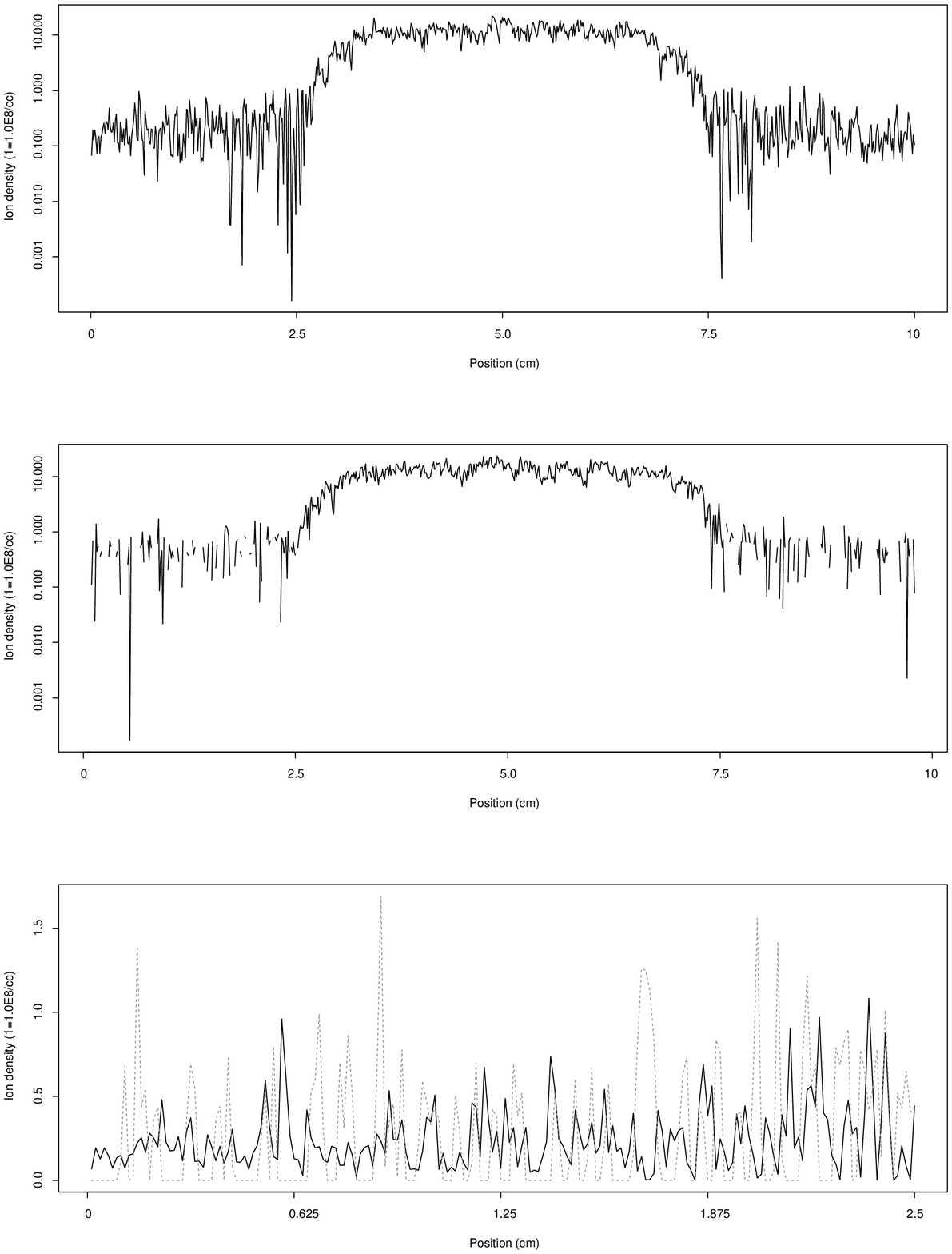,width=\textwidth}
\epsfile{file=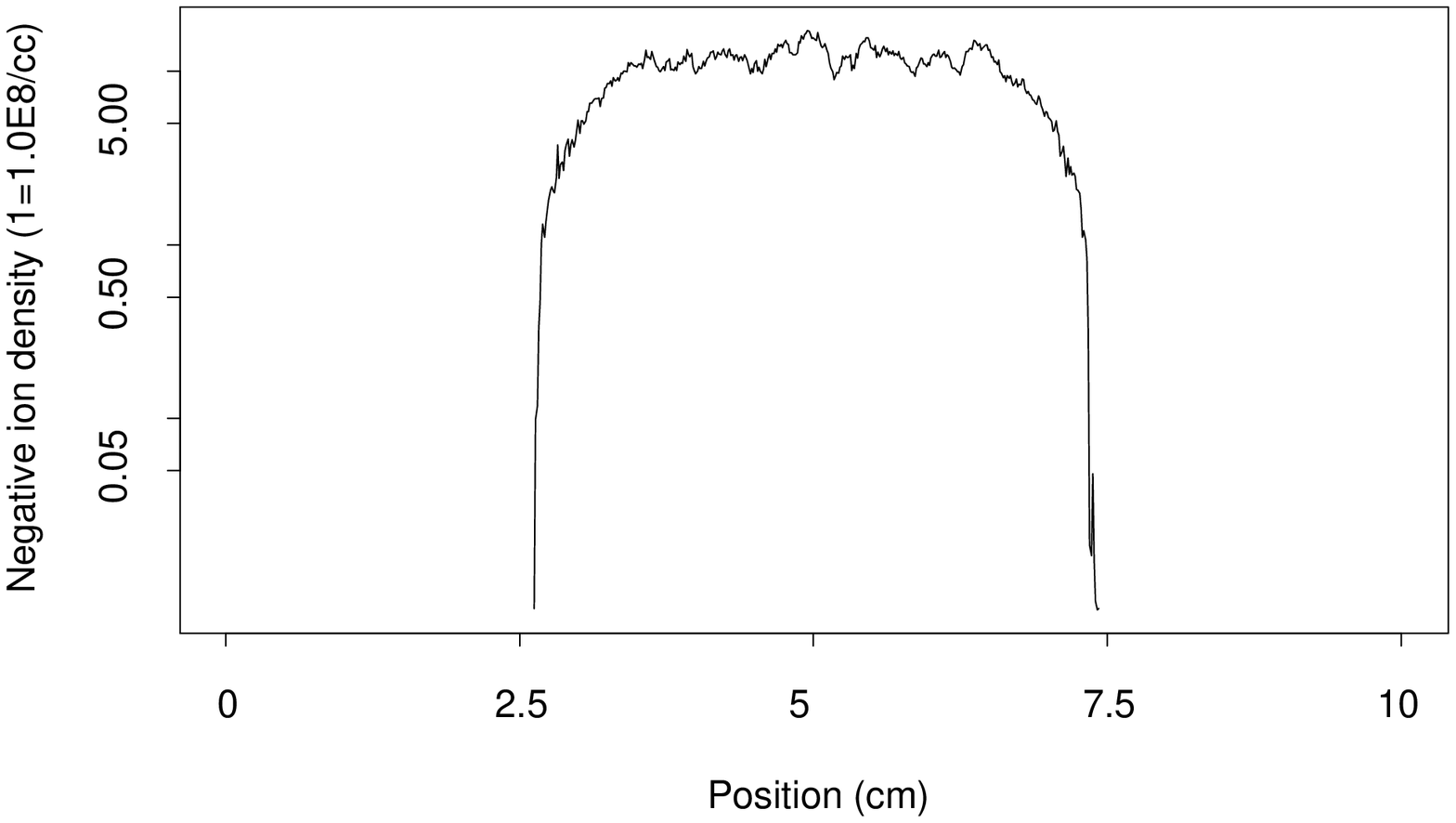,width=\textwidth}
\epsfile{file=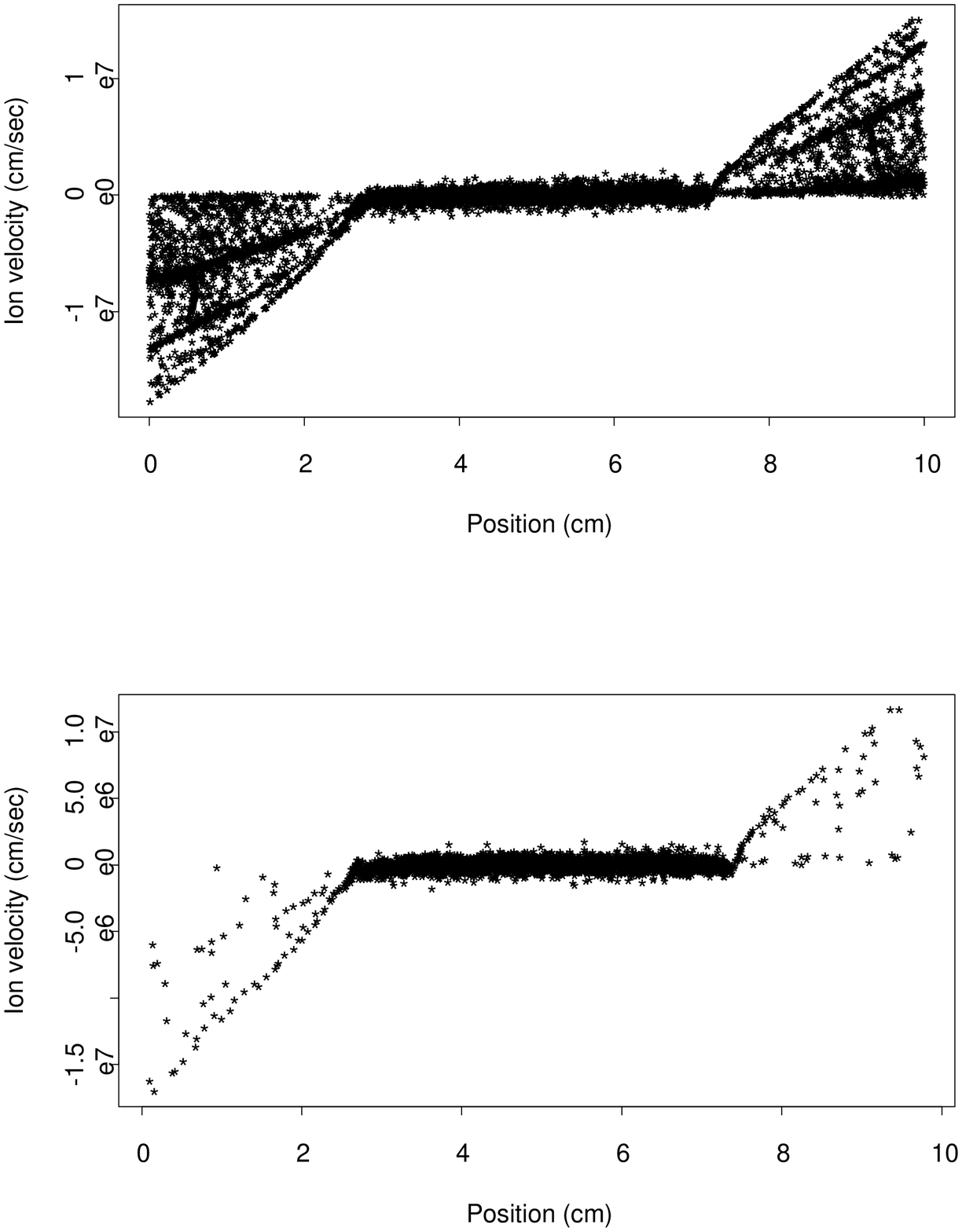,width=\textwidth}
\epsfile{file=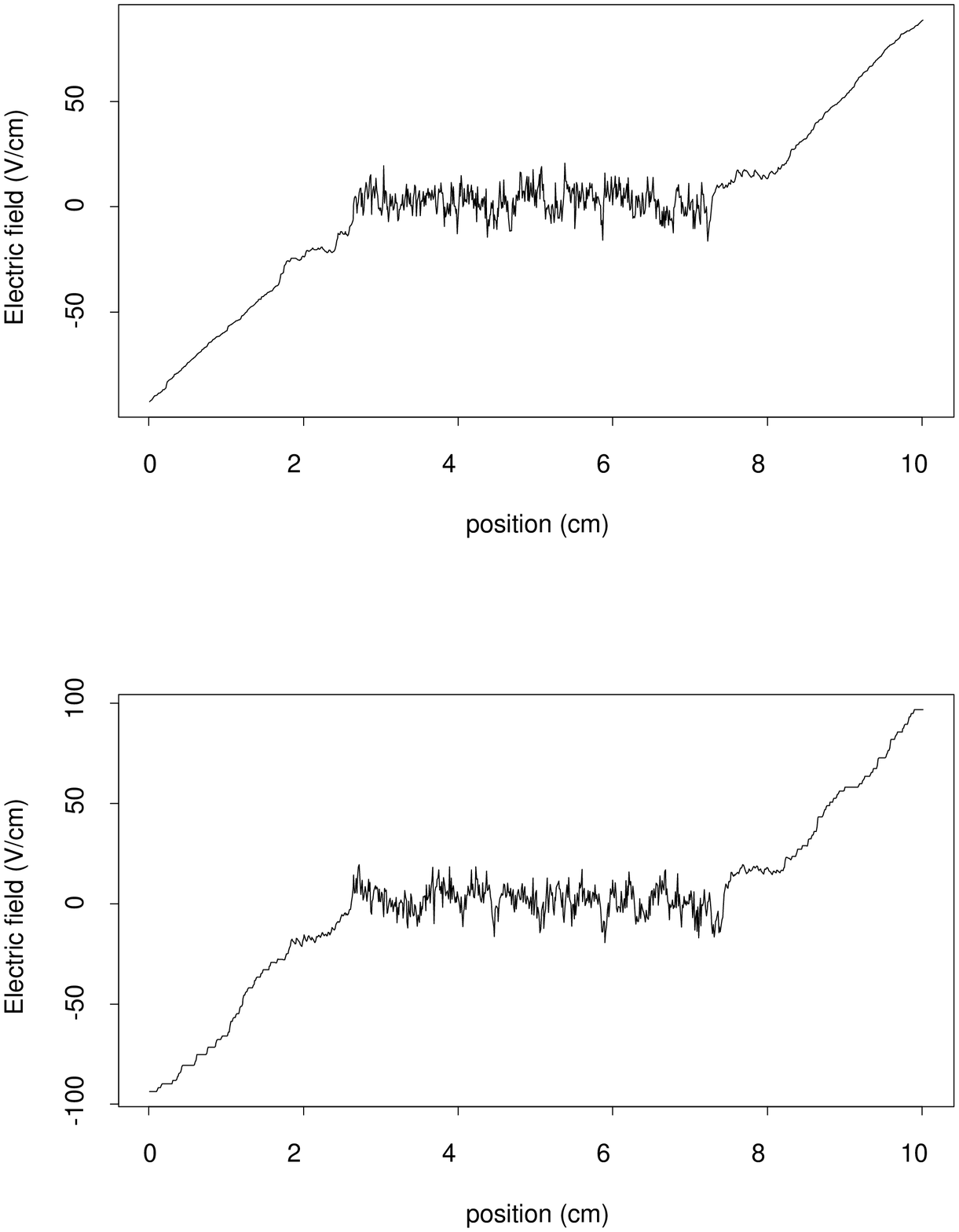,width=\textwidth}
\epsfile{file=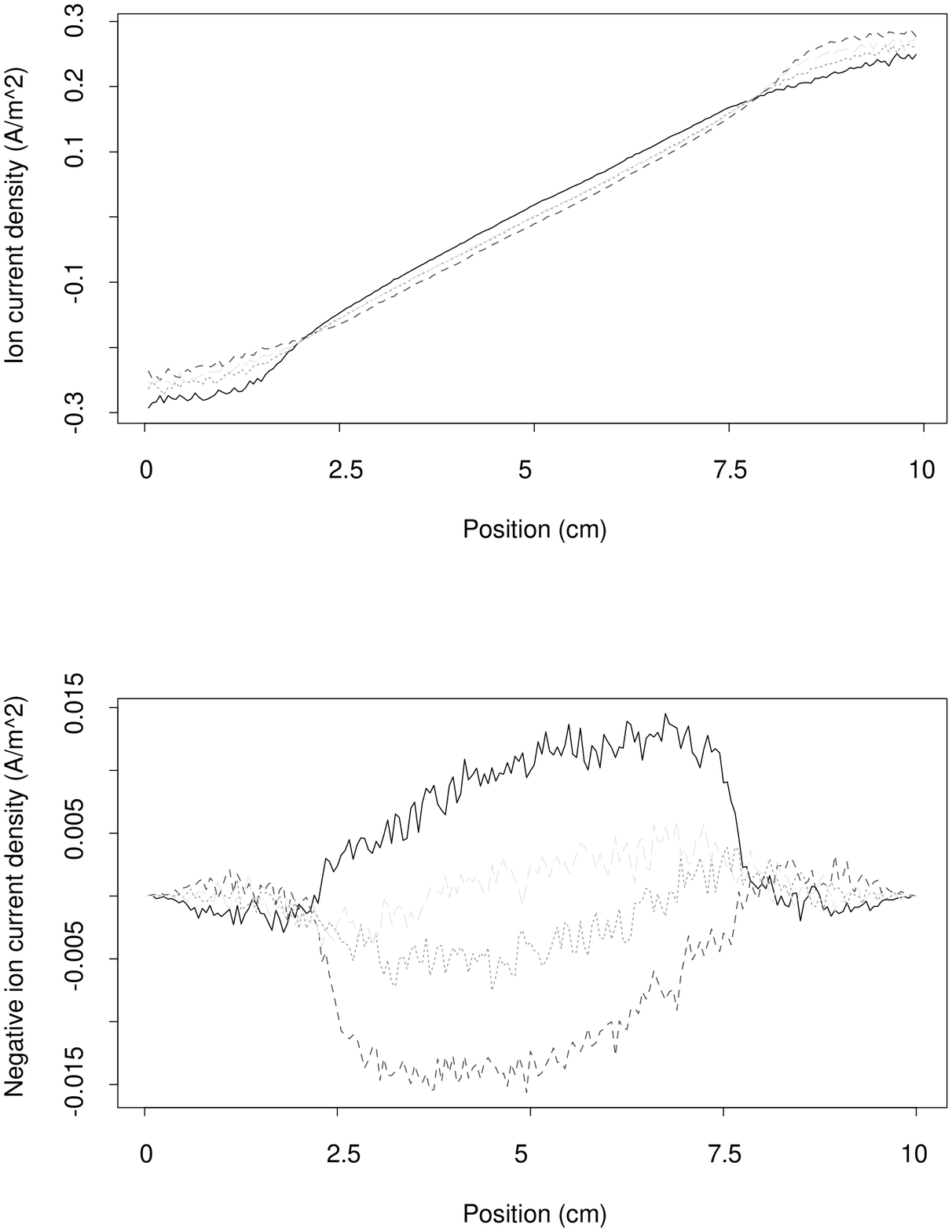,width=\textwidth}
\epsfile{file=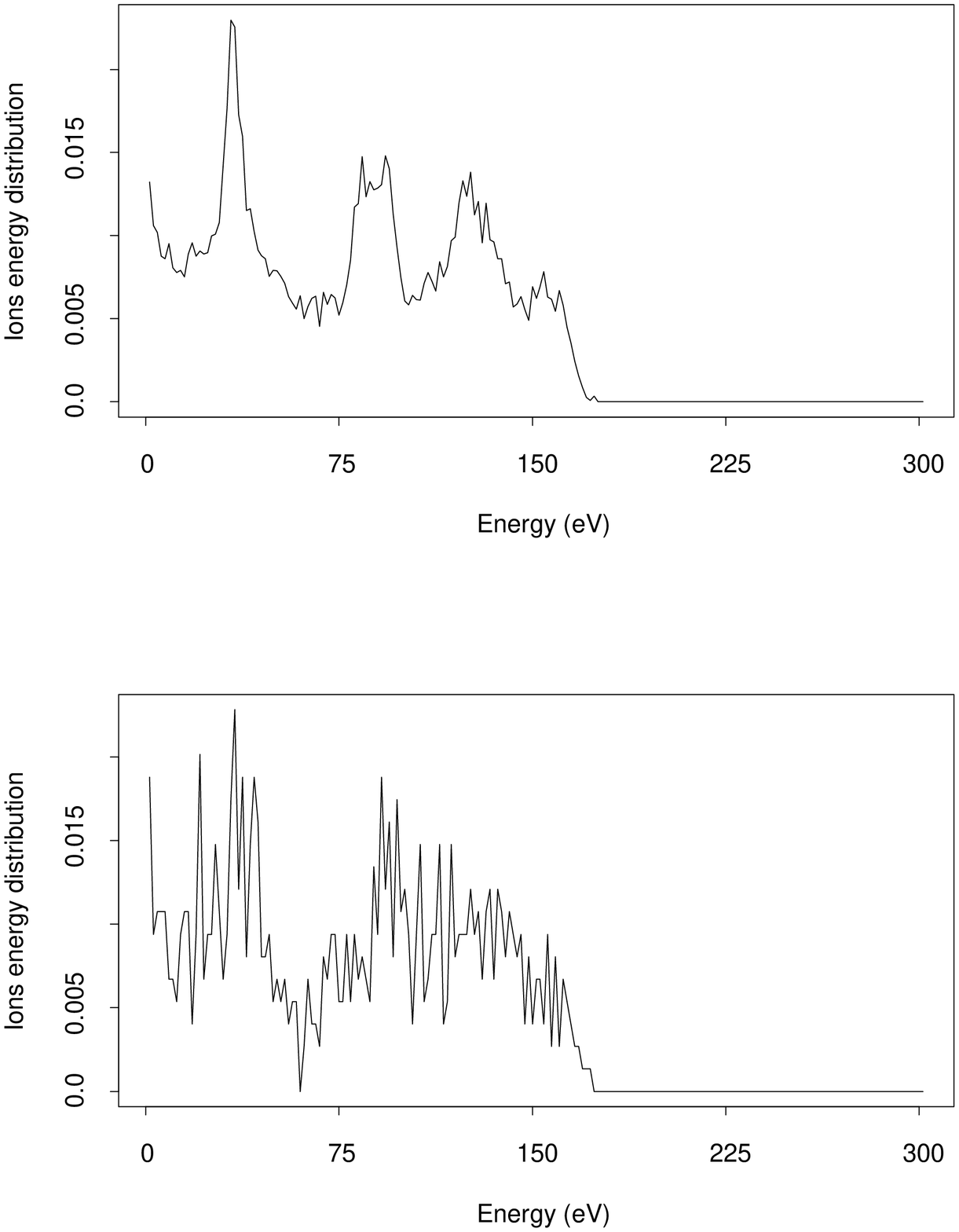,width=\textwidth}
\epsfile{file=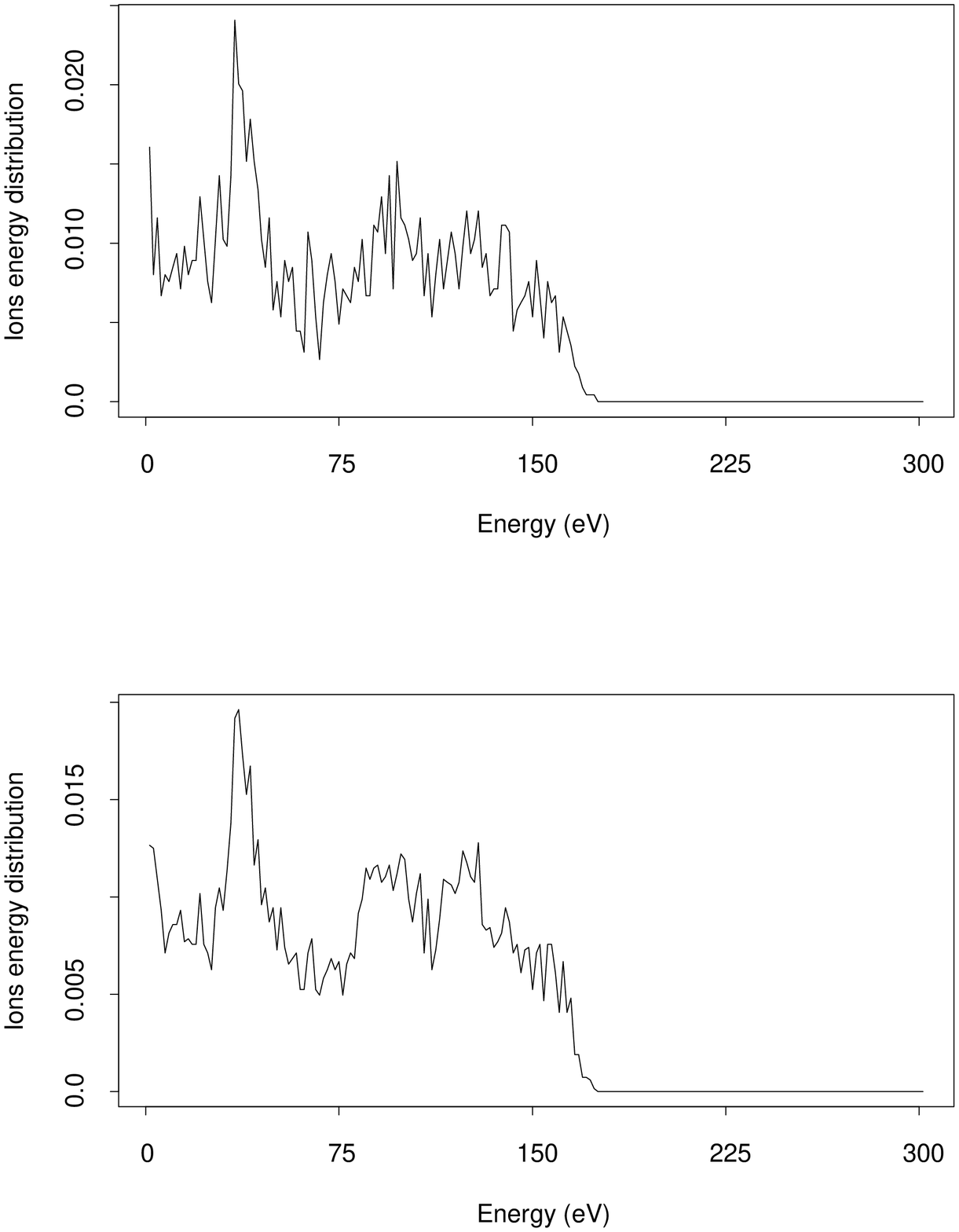,width=\textwidth}

\end{document}